\documentclass[a4paper,fleqn,sort&compress,numbers]{cas-sc}
\usepackage{natbib}
\usepackage{subfiles} 
\usepackage{siunitx}
\usepackage{amsmath}
\usepackage{makecell}
\usepackage{placeins}
\usepackage{booktabs}
\usepackage{multicol}
\usepackage{graphicx}
\usepackage{hhline}
\usepackage{indentfirst}
\def\tsc#1{\csdef{#1}{\textsc{\lowercase{#1}}\xspace}}
\tsc{WGM}
\usepackage{float}

\usepackage{booktabs} 
\usepackage{lineno}
\nolinenumbers

\usepackage{enumitem}

\begin{document}
\let\WriteBookmarks\relax
\def\floatpagepagefraction{1}
\def\textpagefraction{.001}
\shorttitle{Radon emanation system}
\shortauthors{P. Adhikari et al.}

\title [mode = title]{A radon emanation measurement system at the Carleton Noble Liquid Detector Laboratory}
\author[1]{P.~Adhikari}
[
type=editor,
orcid=0000-0002-9395-0560,
]
\ead{pushparajadhikari@cunet.carleton.ca}
\cormark[1]
\author[1]{M.~G.~Boulay}
\author[1]{R.~Crampton}
\author[1]{D.~Gallacher}
\fnmark[2]
\author[1]{M.~Perry}


\address[1]{Department of Physics, Carleton University, Ottawa, K1S 5B6, ON, Canada}

\cortext[cor1]{Corresponding author}
\fntext[2]{Present address: Department of Physics, McGill University, 
3600 University Street, Montreal, QC, H3A 2T8, Canada}

\begin{abstract}
Radon is one of the most important sources of background in rare event search experiments, such as those searching for Dark Matter and neutrinos, due to its unavoidable production from natural uranium. In low-background experiments, radon emanation from detector materials and components accounts for a major portion of contamination. To investigate this, a radon detection system was developed at the Carleton nOble Liquid Detector Laboratory (COLD Lab). The setup consists of a stainless steel emanation chamber, a low-background ZnS(Ag) cell, and an assembly for radon transfer and collection. This setup was used to study radon emanation from materials under vacuum conditions. Additionally, a charcoal trap made of activated charcoal and equipped with a flow meter was constructed to study radon levels in nitrogen gas and the residual radon in the gas filter used in the DEAP-3600 processing system. The radon concentration in the glove box, where critical DEAP-3600 internal detector components were completed, was also calculated based on these measurements. Now calibrated and in-use, the COLD lab radon emanation counter is an essential diagnostic tool for reducing backgrounds in future rate-event search experiments.

\end{abstract}


\maketitle

\section{Introduction}
\label{chap:intro}
\noindent
Radon is an inert noble gas produced through the radioactive decay chains of uranium and thorium, which exist in trace amounts in all materials. Radon is gaseous under standard conditions and can diffuse from deep within materials, often escaping into the spaces between particles by recoiling from its parent during decay. Once trapped within inter-particle spaces, radon will gradually diffuse out of the material through a process known as outgassing. However, only a portion of the generated radon manages to emanate; the rest decays within the solid material before it can escape. The rate at which radon escapes is influenced by several factors, including the material’s microscopic structure, the concentration and distribution of radium, temperature, and moisture content~\cite{nazaroff1988}. Radon from the atmosphere can be trapped inside and on the surface of materials, leading to the build up of out-of-equilibrium Rn decay daughters as shown in the following decay chain, in particular the long-lived isotope $^{210}$Pb with a half-life of 22 years.

\begin{equation}\label{my_chain_eqn}
\begin{aligned}
^{222}\mathrm{Rn} \xrightarrow[\text{\(\alpha\) (5.49 MeV)}]{\text{3.82 d}} ~  ^{218}\mathrm{Po}
\xrightarrow[\text{\(\alpha\) (6.00 MeV)}]{\text{3.05 m}} ~ ^{214}\mathrm{Pb}
\xrightarrow[\text{\(\beta\) (730 keV, max)}]{\text{26.8 m}} ~ ^{214}\mathrm{Bi}
\xrightarrow[\text{\(\beta\) (3.27 MeV, max)}]{\text{19.7 m}} ~ ^{214}\mathrm{Po} 
\xrightarrow[\text{\(\alpha\) (7.69 MeV)}]{\text{0.16 ms}} ~ ^{210}\mathrm{Pb}
\end{aligned}
\end{equation}
Radon supported by internal Radium (Ra) decay is also a significant source of background events in low-background detectors because of its steady emanation rate. The outgassing from Ra and Rn can be clearly distinguished when the emanation times used in the measurements are sufficiently long, due to the significantly shorter half-life of Rn (3.82 days) relative to Ra (1600 years).\\
Rare-event search experiments such as dark matter experiments (DEAP  ~\cite{Amaudruz:2019fx}, \cite{Amaudruz:2018gr}, \cite{DEAPCollaboration:2020hx}, DS-20K \cite{Aalseth:2017um}, \cite{Meyers:2015dc}, LZ~\cite{PhysRevC.102.014602}) and neutrino-less double beta decay (nEXO \cite{Adhikari_2022}, LEGEND-1000 \cite{legend}) must take extreme precautions to avoid radon exposure to sensitive detector components. Deposition of radon daughters onto sensitive detector components is a major source of backgrounds in experiments where expected signals are fewer than one event per year \cite{Ajaj:2019wi}. In dark matter search experiments like DEAP-3600, a significant background source arises from $\alpha$-decays of $^{222}$Rn progeny. One such daughter, $^{210}$Pb, and subsequently $^{210}$Po, may accumulate in the detector, sometimes out of equilibrium with other isotopes in the decay chain. An $\alpha$-particle emitted within the surface of the DEAP-3600 detector can lose energy through interactions with the material, causing it to appear at a lower energy and mimic the expected dark matter signals. In neutrinoless double beta decay experiments such as nEXO, the radon daughter $^{214}$Bi poses a challenge, as its Q-value (3.27 MeV) exceeds the expected $^{136}$Xe double beta decay signal (2.46 MeV). Radon emanating from detector materials can diffuse into the sensitive regions, dissolve in the liquid target, and generate a uniform background of beta decays within the region of interest.\\
Characterizing the radon content and emanation rates of detector materials is a crucial step in the material assay and screening process for sensitive detectors. Elevated radon levels can significantly affect the sensitivity of rare-event search experiments, making it essential to verify and quantify radon concentrations before any material is used in detector construction. For example, the Sudbury Neutrino Observatory (SNO) collaboration measured $^{222}$Rn dissolved in water to monitor and control radon-related backgrounds in their detector~\cite{BLEVIS2004139}. Direct measurements of radon emanation are necessary to accurately determine the expected radon load within detector components. Additionally, estimating radon concentrations within clean environments is important for predicting radon exposure during the fabrication of detector parts in glove boxes, where traditional and commercial radon counters often lack the required sensitivity.


\section{Experimental setup}
\subsection{Radon Emanation Apparatus}
\begin{figure}[]
    \centering
    \includegraphics[width=0.8\linewidth]{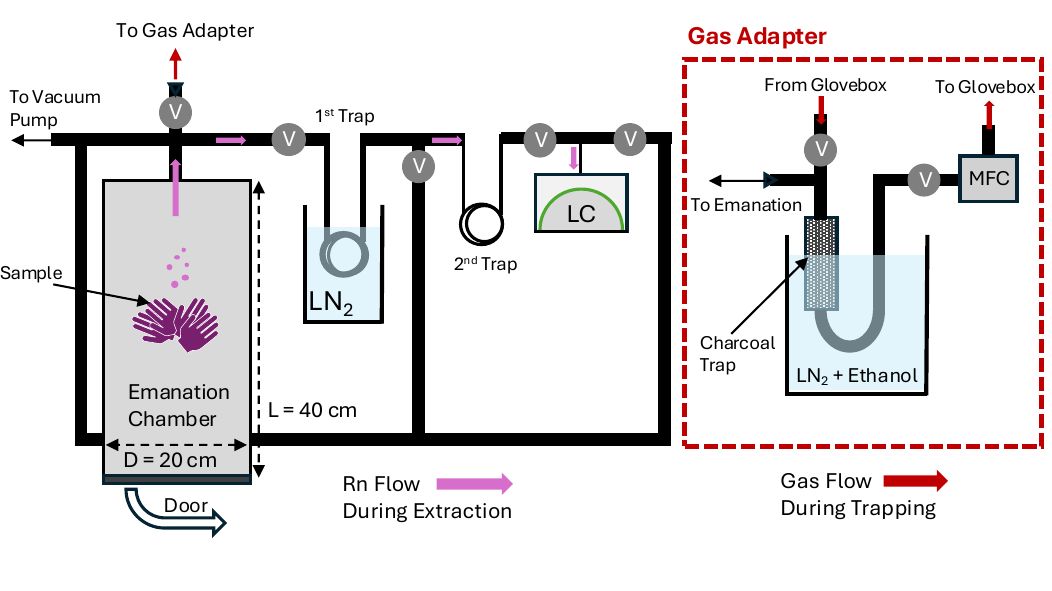}
    \caption{Radon Emanation chamber and gas handling system (left) and optional gas counting adapter (right).}
    \label{fig:emanation}
\end{figure}

\noindent
A cylindrical stainless steel chamber measuring 40 cm in length and 20 cm in diameter was constructed for radon emanation measurements. The chamber is sealed with a ConFlat\textregistered\ flange with copper gasket to ensure vacuum integrity, and a vacuum gauge is installed to monitor the internal pressure. The sample material is enclosed within the chamber, which is connected to an extraction line via a metal valve.\\
 The extraction line includes two radon traps, a primary and a secondary trap, and a vacuum pump. Radon released from the sample is initially collected in the primary trap, which is cryopumped using liquid nitrogen. It is then transferred to the secondary trap, from which it is delivered to a Lucas cell through volume sharing. The volume of the secondary trap is 10\% of that of the Lucas cell. An outline sketch of the system is shown in \autoref{fig:emanation}.\\
 To measure the radon concentration in gas extracted from a compressed gas dewar, an additional adapter is integrated into the system, similar to the setup implemented at the SNOLAB as described in Ref \cite{FATEMIGHOMI2025170422}.  The gas adapter features a charcoal trap submerged in a cooling bath made of a liquid nitrogen–ethanol slurry, which maintains a temperature of $\sim~-$122$^{\circ}$C. At this temperature, activated charcoal effectively adsorbs radon, while allowing the nitrogen gas to flow unimpeded. During measurement, the gas is directed through the charcoal trap to accumulate radon through adsorption. The trap is then connected to the radon collection setup, where the trapped gas is transferred and quantified using the Lucas cell.
\subsection{Lucas cell}
\noindent
A hemispherical Lucas cell, 2 inches in diameter, made of acrylic is designed for the radon collection. The inside of the cell chamber wall is coated with Zinc-Sulphide(ZnS) phosphor as a scintillator, the coating is done in a glovebox filled with nitrogen to reduce Rn load on the scintillation layer.  This is assembled with a quick connector to connect to the radon board. The hemispherical design results in $\sim$33\% light loss; however, the pulse amplitude remains sufficiently large for reliable detection \cite{ManqingLiu1991}.
\begin{figure}[]
    \centering
    \includegraphics[width=0.75\linewidth]{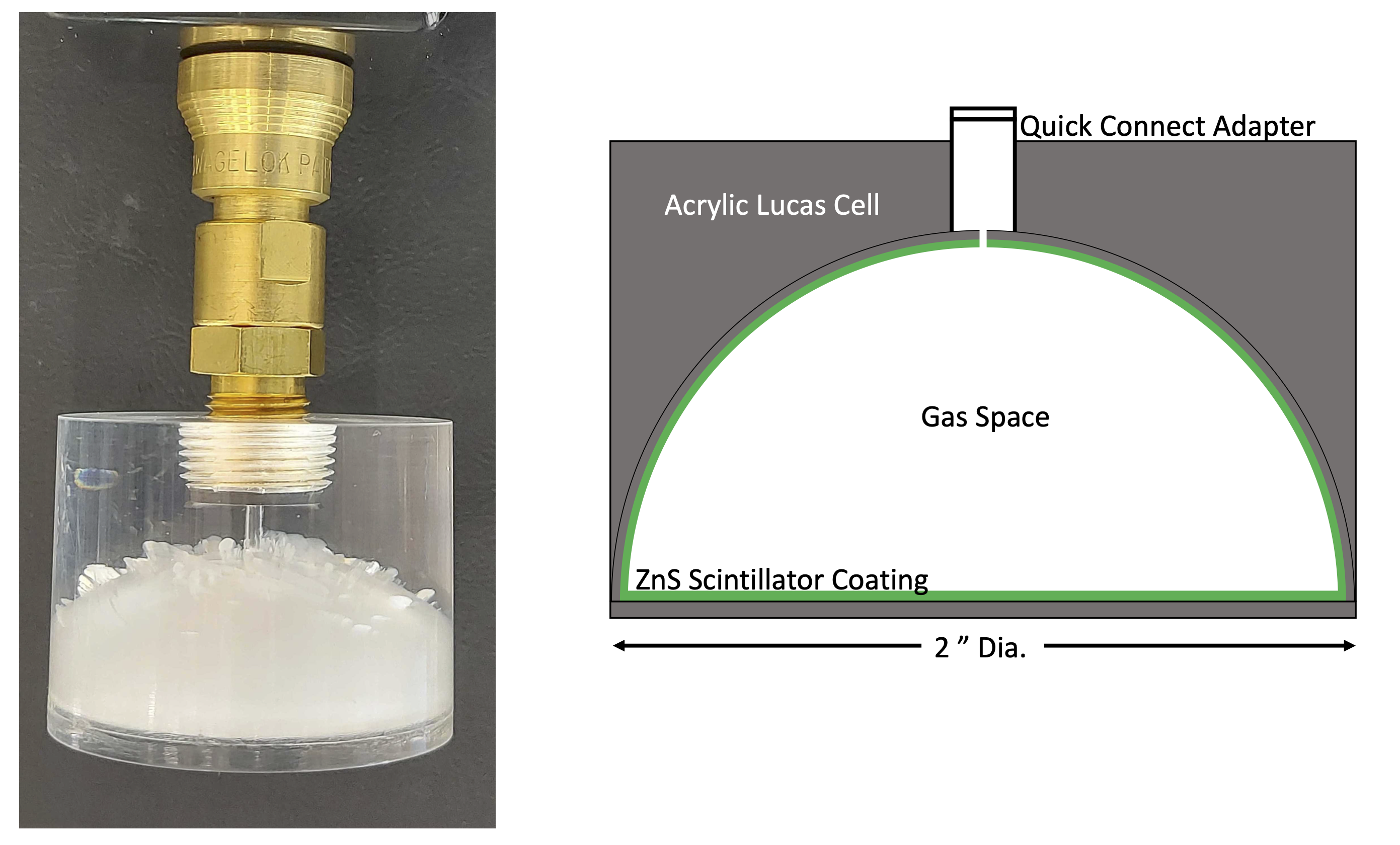}
    \caption{Left: Photo of one Lucas cell prepared at the COLD lab, used for Radon measurements. Right: Labelled cross-section illustration of a Lucas cell, highlighting the coated area for scintillator.}
    \label{fig:lucas_cell}
\end{figure}
\subsection{Operating Procedure}
\noindent
Radon collection and counting involves three main steps: emanation, extraction, and measurement. The procedure is as follows.

\begin{enumerate}
\item Emanation: 
    \begin{enumerate}[label=\alph*)]
    \item Back-fill the chamber with liquid nitrogen boil-off gas when bringing the system up to atmospheric pressure.
      \item Open the chamber and place the sample inside the chamber and close it.
      \item Pump on the chamber for 24 hours and with all valves in the system closed.
    \end{enumerate} 

\item Extraction:
    \begin{enumerate}[label=\alph*)]
      \item Open all valves except for valves from the chamber and pump the radon board for at least 2 hours.
      \item Isolate the secondary trap, by closing a valve between primary and secondary trap, and connect the Lucas cell to the quick connect port.
      \item Cool the large trap with liquid nitrogen and begin the extraction by slowly opening the chamber valve, and continue extraction for 1 hour. 
      \item Isolate the primary trap from the chamber and warm the large trap with heat gun.  
      \item Cool the secondary trap and pump the large trap through the secondary trap for 10 minutes.
      \item Close all valves, stop pumping, warm the secondary trap and transfer the gas from the secondary trap to Lucas cell.
    \end{enumerate}

\item Measurement: 
    \begin{enumerate}[label=\alph*)]
      \item Remove the Lucas cell from the apparatus, cap the Lucas cell, then place in the counting slot on the PMT.
      \item Run the DAQ and take data for at least 24 hours.
    \end{enumerate}
\end{enumerate}

\subsection{Data Acquisition}
\begin{figure}[]
    \centering
    \includegraphics[width=0.9\linewidth]{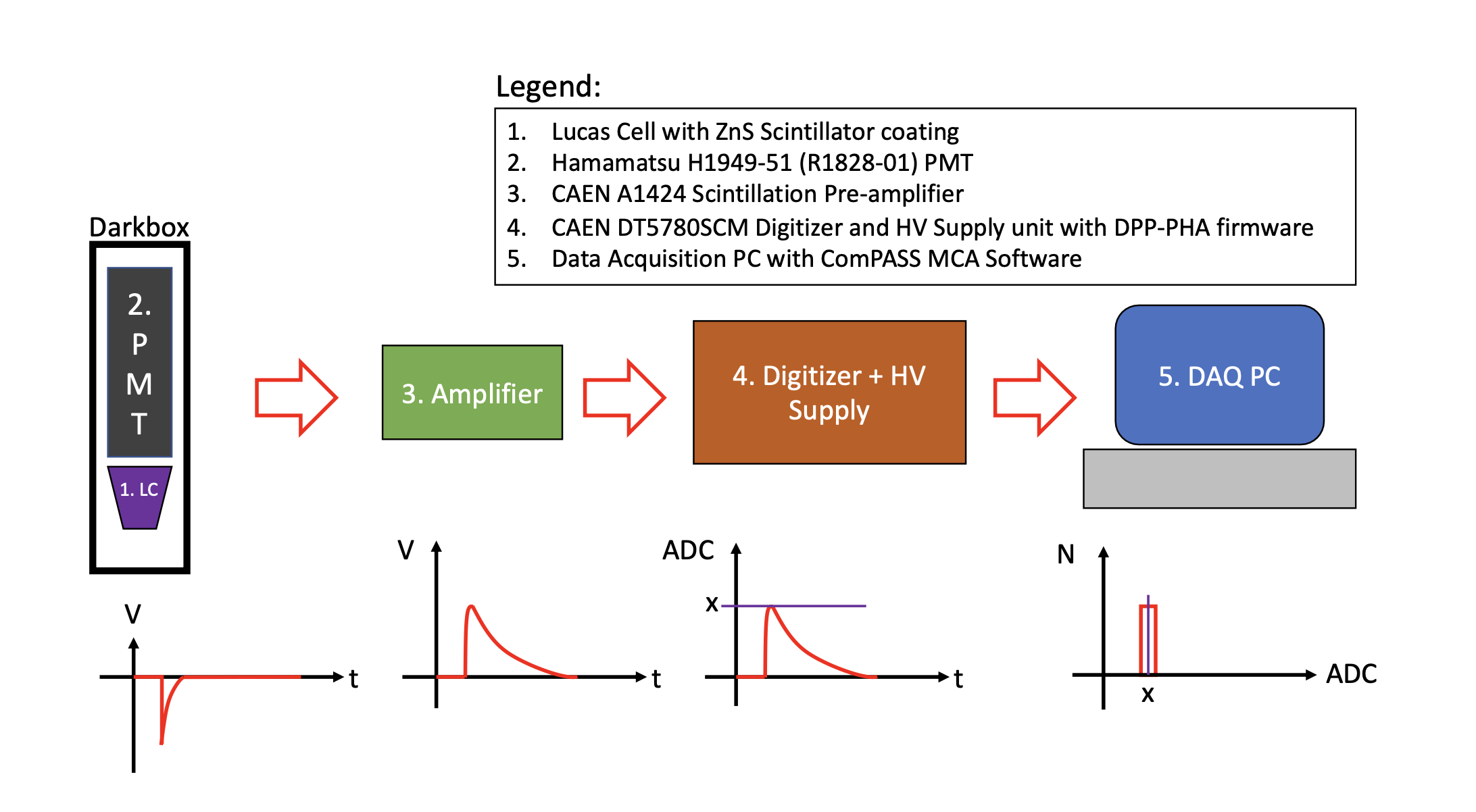}
    \caption{Schematic of DAQ system for Radon emanation setup at Carleton. PMT signals are amplified by the CAEN pre-amplifier, and digitized by the DT5780SCM, with automatic pulse-height analysis firmware. Data is recorded by the DAQ PC running ComPASS software for subsequent analysis.}
    \label{fig:daq}
\end{figure}

\noindent
The data acquisition system (DAQ) for the COLD Lab radon assay setup is a straightforward system.  $\alpha$-decays from radon progeny produce scintillation light in ZnS which is peaked at $\sim$450 nm \cite{article_ZnS}. This light is transmitted through the transparent acrylic to a 2" Hamamatsu R1828-01 photo-multiplier tube (PMT) (Inside the H1949-51 package) inside a dark box. The PMT signals are amplified by a CAEN A1424 scintillation charge-sensitive pre-amplifier which inverts the signal, and performs charge integration. The amplitude of the output pulse from the pre-amplifier is then proportional to the charge in the PMT pulse. The amplifier signal is sent to a CAEN DT5780SCM desktop digitizer, which contains a 14-bit 100 MS/s ADC, and provides the HV for the PMT.CAEN DPP-PHA firmware installed on the digitizer performs live pulse-analysis, extracting the time and height in ADC counts of incoming signals \cite{website_dpp_pha}. The pulse data, as well as run information and dead-time estimates, are transmitted to the DAQ PC over a USB link and saved to a ROOT \cite{website_root} file by the ComPASS software package from CAEN. Analysis is performed on the output ROOT file, using custom analysis scripts. The signal processing and DAQ components are shown in \autoref{fig:daq} for clarity.


\section{Analysis}
\subsection{Calibration and Background measurement}
\begin{figure*}[]
  \begin{center}
      \includegraphics[width=0.5\textwidth]{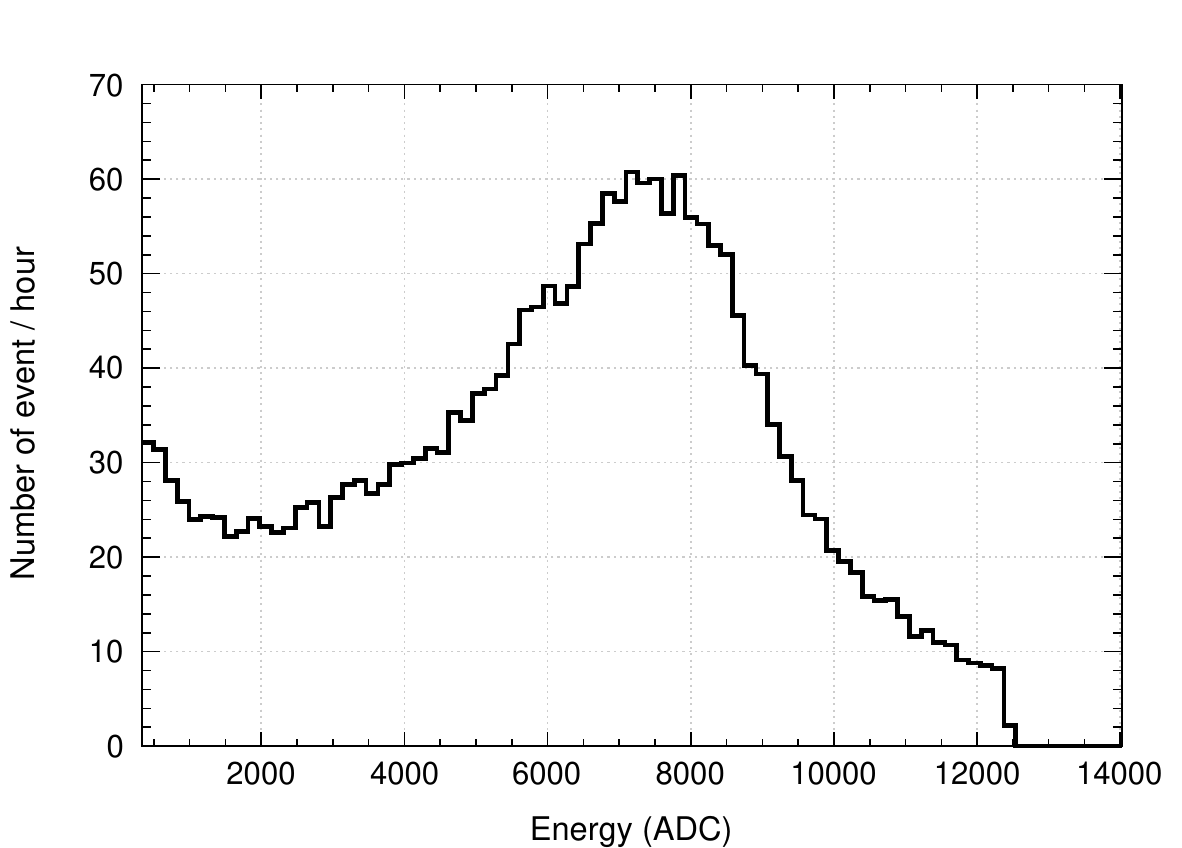}
  \end{center}
  \caption{Alpha energy spectrum from the Buna gasket sample}
  \label{ref:dust_sim}
\end{figure*}

\noindent
For system calibration, a radon source with a well-defined alpha energy peak is required. To achieve this, the Lucas cell was filled with radon emanated from a Buna rubber panel, which is known to have a high radon emanation rate exceeding 30,000 atoms m$^{-2}$h$^{-1}$. Due to the short half-life of $^{214}$Po, as illustrated in Equation \ref{my_chain_eqn}, the dominant alpha emissions observed during the initial hours are primarily from $^{214}$Po decay, which produces alpha particles with an energies of 7.69 MeV. This leads to a broad spectral feature that serves as the primary energy calibration reference, as shown in Figure \ref{ref:dust_sim}. Due to the Lucas cell’s limited energy resolution, no additional discrete peaks are distinctly visible. However, an increasing number of low-amplitude events are observed below 1500 ADC counts, mainly due to electronic noise. To suppress pile-up and noise contributions, we apply a minimum signal amplitude threshold of 1500 ADC counts.\\
Before taking sample measurements, it is essential to determine both the background and efficiency of the detection system. Two primary sources of background are identified. The first originates from intrinsic activity within the Lucas cell, primarily due to radioactivity in the ZnS scintillation powder and the presence of  $^{210}$Po deposited on the acrylic surface of the cell. This background is quantified by operating the Lucas cell under vacuum conditions, where approximately five alpha events are observed per day.\\
The second source of background arises from radon emanation associated with the radon board and the emanation chamber. To estimate the total background contribution from the entire system, radon is extracted from an empty chamber. The empty chamber is evacuated for 24 hours, sealed for an additional 24 hours, and then sampled for radon. This measurement yields approximately seven alpha events per day, representing the total system background, including the contribution from the Lucas cell.\\




\subsection{Efficiency}
\begin{figure*}[]
  \begin{center}
      \includegraphics[width=0.5\textwidth]{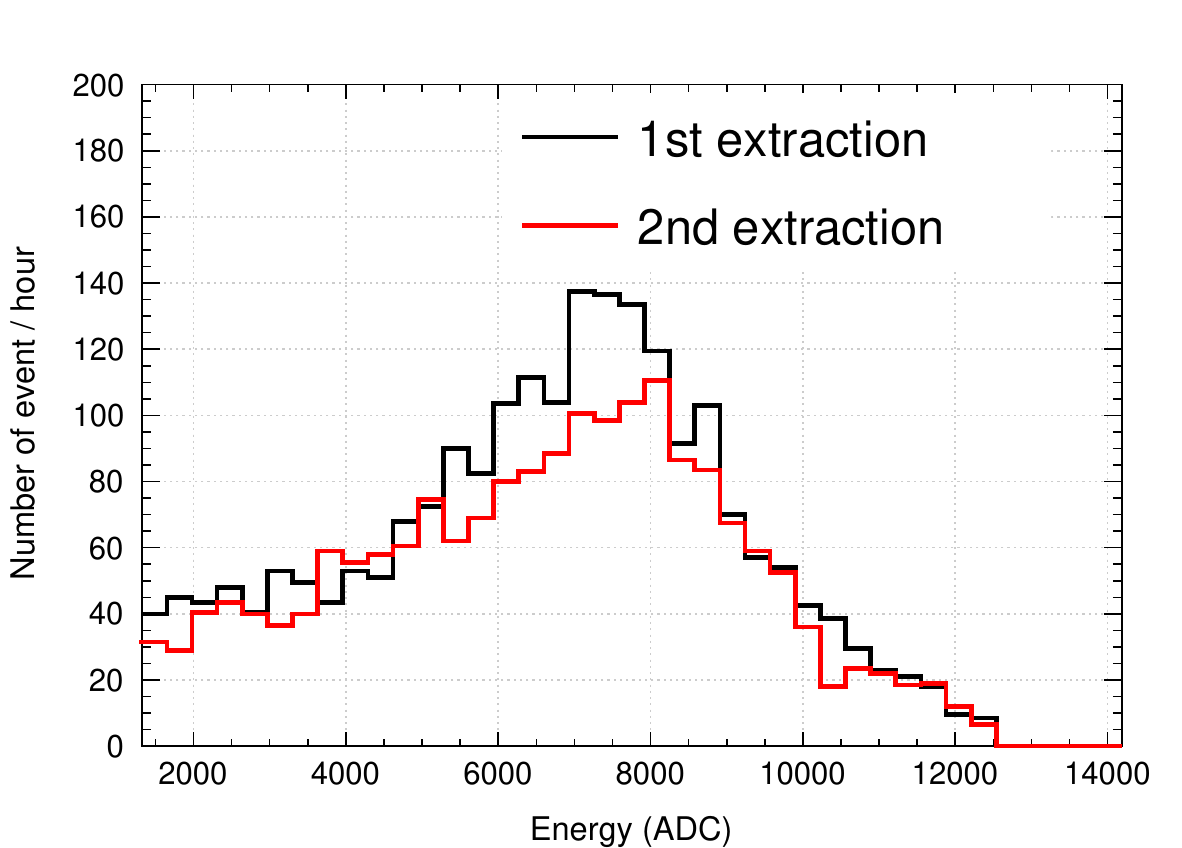}
  \end{center}
  \caption{Comparison of the alpha energy spectra obtained from the Buna gasket sample. The black curve represents the spectrum from the initial extraction, while the red curve corresponds to the second extraction. }
  \label{ref:eff}
\end{figure*}

\noindent
The overall efficiency of the radon measurement system is quantified in two distinct components: the collection of emanated radon within the Lucas cell (collection efficiency) and the detection of alpha particles as determined by the geometric configuration of the Lucas cell (detection efficiency). The collection efficiency is assessed using a radon source derived from a Buna-N gasket. The methodology for determining the collection efficiency is:

\begin{enumerate}
  \item Collect radon from the source into the chamber and record the number of alpha particles detected, $N_{\alpha}$.
    \item Connect the Lucas cell to the chamber and allow 20 minutes for equilibration. Since the chamber is approximately 1,000 times larger than the Lucas cell, the majority of radon will remain in the chamber.
     \item Measure the radon activity in the Lucas cell to determine any residual radon present $N^{r}_{Rn}$.
      \item Transfer the radon from the chamber back into the Lucas cell and record the corresponding alpha counts $N_{Rn}$.
     \item Apply a decay-time correction to the initial alpha count to account for the approximately 5-hour delay between the first and second extractions, yielding the corrected value $\hat{N_{\alpha}}$. Where $\hat{N_{\alpha}} = N_{\alpha}e^{-\lambda\Delta t}$, for $\Delta t\approx$~5~h.
\end{enumerate} 
The efficiency of the system is calculated using Equation \ref{eq:eff}, to be 89\%.
\begin{equation}
\label{eq:eff}
 \epsilon = \frac{N^{r}_{Rn}-N_{Rn}}{\hat{N_{\alpha}}}
\end{equation}
The gas extracted from both the chamber and the primary trap is recovered at 100\% efficiency. Therefore, the observed 11\% reduction in overall efficiency is primarily attributed to the shared volume between the secondary trap and the Lucas cell, as the volume of the secondary trap is approximately 10\% of that of the Lucas cell. Figure \ref{ref:eff} presents the alpha spectrum from the second extraction, compared to that of the initial extraction, conducted after a 5-hour interval.\\
The detection efficiency of the same dimension Lucas cell was measured at Queen's University, ~\cite{ManqingLiu1991}, which is 63\%. So, the total efficiency of the system is,
\[\epsilon = 0.63 \times 0.89 = 56\%\]


\subsection{Material Radon Assay Results}
\begin{figure*}[]
  \begin{center}
      \includegraphics[width=0.5\textwidth]{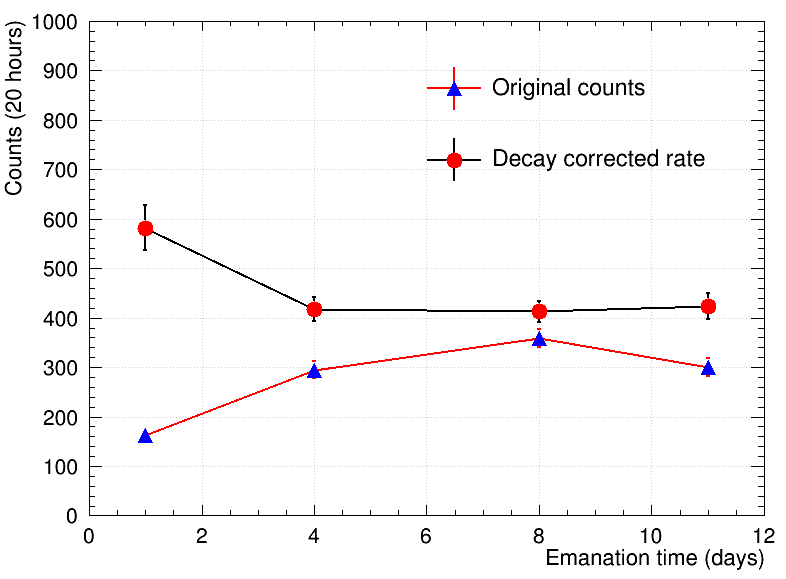}
  \end{center}
  \caption{Radon emanation rates were measured for the two butyl gloves used in the glove box. Early measurements were excluded from the analysis, as they reflect radon outgassing from inter-particle spaces within the material, leading to elevated rates. The later data points, representing the steady-state emanation, were used to determine the radon emanation rate of the material.}
  \label{ref:rate1}
\end{figure*}

\noindent
The majority of the samples measured in the system were intended for the DEAP-3600 upgrade and for preparation for future low-background experiments. The total background of the system is $\sim$~7 events per day, which was subtracted from the sample measurement data. The number of alpha particles detected from the radon collected in the Lucas cell is counted over a 24-hour period; this count can then be converted to the number of radon atoms emanated from the surface of the materials. Since radon undergoes two subsequent alpha decays, the relation in Equation \ref{eq:emanation} can be used to convert the measured alpha count into the total number of emanated radon atoms,
\begin{equation}
\label{eq:emanation}
N_{Rn,Em} = \frac{N_{\alpha}}{(1-e^{(\lambda/\tau)\cdot t)}) \cdot 3 \epsilon} . 
\end{equation}
In Equation \ref{eq:emanation}, $N_{Rn,Em}$ is the estimated number of radon atoms emanated from the sample. As input, $N_{\alpha}$ is the number of alpha events observed by the counter, $\lambda$ is the Rn decay constant, $\tau$ is the half-life of radon, $\epsilon$ is the total efficiency of the radon emanation system and the emanation time for the sample is $t$. \\
Data were collected over several weeks through multiple extractions and plotted as emanation time versus counts, as shown in Figure \ref{ref:rate1}. Early measurements are excluded from the analysis because they reflect radon outgassing from inter-particle spaces within the material, resulting in elevated rates. Radon supported by internal radium (Ra) decay exhibits a steady emanation rate; therefore, measurements should be taken after the rate stabilizes. The later data points in Figure \ref{ref:rate1} represent this steady state and were used to estimate the radon emanation rate from the material samples. The results of the material assays are summarized in Table \ref{tab:assay_results}.

\begin{table*}[H]
    \centering
        \caption{Material assay results}

    \begin{tabular}{|p{0.167\linewidth}|p{0.167\linewidth}|p{0.267\linewidth}|}

      \hline
      \textbf{Material Name}    & \textbf{Sample Quantity}  & \textbf{Emanation Rate (Rn atoms/h) } \\ 
      \hline
      Silicone gasket  & 1012.46 cm$^{2}$ &  141.2~$\pm$~28 per m$^{2}$\\
      \hline
     Butyl gasket  & 712.1 cm$^{2}$	 & 186.5~$\pm$~37 per m$^{2}$ \\
      \hline
     Buna-N gasket  & 712.1 cm$^{2}$	 & 31714.6~$\pm$~1400 per m$^{2}$\\
      \hline
     Butyl glove  &2 gloves & 10~$\pm$~2 per glove \\
      \hline
      EPDM O-ring  & 588 cm & 11~$\pm$~2 per m \\
      \hline
    \end{tabular}
    \label{tab:assay_results}
\end{table*}

\subsection{Steady-state Radon Numerical Analysis}
\noindent
A glove box purged with a continuous flow of liquid-nitrogen boil-off gas was used for sanding and coating the acrylic flow guide. To estimate the total radon load within this environment, a detailed accounting of all radon-emitting components inside the glove box was performed. In addition to radon outgassing from internal materials, two additional contributions were considered in calculating the steady-state radon concentration: (1) residual radon introduced from room air and impure nitrogen prior to the transition to boil-off nitrogen, and (2) the intrinsic radon content of the boil-off nitrogen itself.\\
The steady-state radon concentration inside the glovebox can be described by \autoref{eq:steadystate}.
\begin{equation}
    \label{eq:steadystate}
    a(t) = a_{LN2}(t) + a_{res}(t) + a_{emanation}(t) \; \mathrm{[Bq_{Rn222}/m^{3}]}
\end{equation}
Where, $a_{X}$ represents the specific activity of $^{222}$Rn from component `X'. The dominant contributions to the total emanation rate are shown in \autoref{tab:steadystate_eman}, measured in the system described here. \\
\begin{table}[]
    \centering
    \caption{Dominant radon emanation components in the glove-box for steady-state analysis, components under 1 atom/h not listed. These components are directly measured by emanating the individual parts in the vacuum chamber.}
    \begin{tabular}{|l|c|}
        \hline
        \textbf{Material} & \textbf{Radon Emanation Rate (Atoms/h)} \\
        \hline
        EPDM Window Gasket x2 & 170  \\
        \hline
        EPDM O-rings & 20 \\
        \hline
        Gloves x4 & 20 \\
        \hline
        Aluminum Components & 5.5 \\
        \hline
        Total & 215.5 \\
        \hline
    \end{tabular}
    \label{tab:steadystate_eman}
\end{table}
The liquid nitrogen boil-off gas is purged through the glovebox at a rate of $\sim\,$14 L/min, or $\sim\,$1 volume turnover per hour for the $\sim\,$0.85 m$^{3}$ glovebox. The steady-state of radon inside the glovebox from \autoref{eq:steadystate}, as a sum of the individual contributions, taking into account the flow-through of nitrogen gas. The mean lifetime of nitrogen gas in the glovebox is described by \autoref{eq:tau_ln2}, and is taken into account in the specific activity contributions. 
\begin{equation}
    \label{eq:tau_ln2}
    \tau_{\text{flow}} = \frac{V_{\text{glovebox}}}{Q}
\end{equation}
Where `Q' is the volumetric flow rate, in glovebox volumes per hour. 
\begin{equation}
    \label{eq:LN_2} 
    a_{LN2}(t) = a^{0}_{LN2} \times \left(1.0 - e^{-t/\tau_{flow}}\right)
\end{equation}

\begin{equation}
    \label{eq:res}
    a_{res}(t) = a_{room}\times \left(e^{-(t-t_{0})/\tau_{flow}}\times e^{-(t-t_{0})/\tau_{Rn_{222}}}\right) 
\end{equation}
Where '$\tau_{Rn_{222}}$' is the mean lifetime constant of $^{222}$Rn, is 132.4 hours, and '$a_{room}$' is the activity of Rn222 in the room, measured to be $\sim\,$3 Bq/$m^{3}$ with a Radon meter. \\
We estimate each emanated radon contribution to the steady-state radon concentration by solving the differential equation shown in \autoref{eq:diff_em}. \\
\begin{equation}
    \label{eq:diff_em}
    \frac{dN_{i}}{dt}(t) = R^{i}_{Rn_{222}} - N(t)\times \left( \frac{1}{\tau_{flow}} + \frac{1}{\tau_{Rn_{222}}} \right)
\end{equation}
Where `$R^{i}_{Rn_{222}}$' is the emanation rate for each `i-th' component inside the glovebox. The solution of \autoref{eq:diff_em} gives the contribution to the total radon concentration from each component and is outlined in \autoref{eq:eman}.\\
\begin{equation}
    \label{eq:eman}
    a_{\text{emanation}}(t) = \sum\limits_{i}^{n}\,\frac{R^{i}_{Rn222}\ln{(2)}}{3600\, \tau_{Rn_{222}}} \times \frac{\tau_{\text{flow}}\tau_{Rn222}}{\tau_{\text{flow}} + \tau_{Rn222}} \times \left( 1.0 - e^{-(t-t_{0})/\tau_{\text{flow}}}\times e^{-(t-t_{0})/\tau_{Rn222}} \right)
\end{equation}
Where, all `n' contributions shown in \autoref{tab:steadystate_eman}, as well as any additional components not listed, which have a total radon emanation load of $\sim$~10 atoms/h, are summed together to estimate the total radon concentration due to emanation as a function of time. 

\subsection{Gas Counting Results}
\noindent
Radon concentration in the nitrogen gas supplied from the Dewar is measured by integrating the radon emanation system with a charcoal trap, as illustrated in Figure \ref{fig:emanation}. At cryogenic temperatures, activated charcoal captures radon, which is subsequently released by heating. To collect radon, the charcoal trap is immersed in liquid nitrogen-ethanol slurry while nitrogen gas is passed through it. The trap is connected to the radon emanation chamber, and the collected radon is then transferred to a Lucas cell for counting.\\
This setup is also employed to measure radon originating from the glove box by circulating nitrogen gas through the enclosure. Gaseous nitrogen enters the glove box and exits through an outlet connected to the charcoal trap. The radon collected in this process reflects both the radon content of the nitrogen gas and the emanation from the glove box itself. The counting results, summarized in Table \ref{tab:n2assay}, show that measurements of individual components inside the glove box (listed in \autoref{tab:steadystate_eman}) are consistent with the overall gas assay results.
\begin{table*}[h!]
    \caption{Nitrogen gas assay results using a charcoal trap in the gas outlet of the glove-box. The gas assay results are consistent with the values estimated with the individual parts measurement in the emanation chamber.}
    \centering
    \begin{tabular}{|p{0.167\linewidth}|p{0.09\linewidth}|p{0.06\linewidth}|p{0.07\linewidth}|p{0.2\linewidth}|p{0.15\linewidth}|}
      \hline
      \textbf{Nitrogen Gas Flow} & \textbf{Flow Rate (LPM)} & \textbf{Time (h)} & \textbf{Counts (20 h) }  & \textbf{Alpha Rate ($\mu$Bq/m$^{3}$) ($\epsilon$ and bkgd. corrected)}  & \textbf{Radon Emanation Rate in Glove Box (atoms/h)} \\
      \hline
      Dewar & 2.4 & 7 & 61 &1264~$\pm$~200 &\\
      \hline
      Dewar & 2.4 & 7 & 40 & 744~$\pm$~185& \\
      \hline
      Dewar and glove box & 2.4 & 7 & 209  &4960~$\pm$~760 & 178~$\pm$~27 \\
      \hline
      Dewar and glove box & 2.4 & 7 & 237  & 5630~$\pm$~844 & 203~$\pm$~30 \\
      \hline
    \end{tabular}
    \label{tab:n2assay}
\end{table*}



\section{Discussion}
\label{chap:Sec6}
\noindent
Radon, produced from trace amounts of natural uranium, is one of the most significant sources of background in rare-event searches, such as dark matter and neutrino experiments. In low-background detectors, radon emanation from construction materials and components is a major contributor to contamination. To address this, a radon detection system was developed, consisting of a stainless steel emanation chamber, a low-background ZnS(Ag) scintillation cell, and an assembly for radon transfer and collection. This setup enables the study of radon emanation from materials under vacuum conditions.\\
In addition, a charcoal trap filled with activated charcoal and equipped with a flow meter was constructed to measure radon levels in nitrogen gas and to quantify residual radon in the gas filter used in the DEAP-3600 purification system. Using this system, the concentration of radon in the glovebox in which the DEAP-3600 internal detector components are machined, sanded, and coated was also determined. Measurements of individual glove box components in the vacuum chamber were found to be consistent with those obtained using gas-flow sampling through the charcoal trap connected to the glove box outlet.\\
This system is now fully operational and characterized, and will also be employed to measure radon emanation from materials used in the DarkSide experiment ~\cite{2024_DS}. This will include an assessment of radon content in the materials used for the TPB coating evaporation chamber for the DarkSide acrylic panels. We are planning upgrades to the system  to allow for temperatur regulation within the vacuum chamber in order to study the temperature dependence of emanation rates for critical detector components. By doing so we can further optimize construction and assembly proceedures to reduce radon load for future ultra-low background detectors.


\subsection*{Acknowledgements}
\noindent
We thank the Natural Sciences and Engineering Research Council of Canada, the Canadian Foundationfor Innovation (CFI), the Ontario Ministry of Research and Innovation (MRI), Carleton University, the Canada First Research Excellence Fund, and the Arthur B. McDonald Canadian Astroparticle Research Institute.

\appendix

\printcredits

\bibliographystyle{elsarticle-num}
\bibliography{refs}

\end{document}